\documentclass[12pt]{iopart}
\usepackage{epsf}  
\begin{document}

\title{
The Horn and the Thermal Model.
}
\author{
{\sc J. Cleymans$^a$,
H. Oeschler$^b$,
K. Redlich$^c$,
S. Wheaton$^a$}
}
\address{
$^a$ UCT-CERN Research Centre and  Department of Physics, University of Cape Town,
Rondebosch 7701, Cape Town, South Africa\\
$^b$ Institut f\"ur Kernphysik,
Technische Universit\"at Darmstadt, D-64289~Darmstadt, Germany,\\
$^c$ Institute of Theoretical Physics, University of Wroc\l aw,
 Pl-45204 Wroc\l aw, Poland 
}

\begin{abstract} 
The recently discovered sharp peak in the $K^+/\pi^+$ ratio
in relativistic heavy-ion collisions is
discussed in the framework of the thermal model. 
In this model
a rapid change is expected as the hadronic gas undergoes a
transition from a baryon-dominated  to a meson-dominated gas. The
transition occurs at a temperature $T$ = 140 MeV and baryon
chemical potential $\mu_B$ = 410 MeV corresponding to an incident
energy of $\sqrt{s_{NN}}$ = 8.2 GeV. 
\end{abstract}
\begin{figure}[ht]
\centerline{
\epsfysize=32pc
\epsfxsize=24pc
\epsfbox{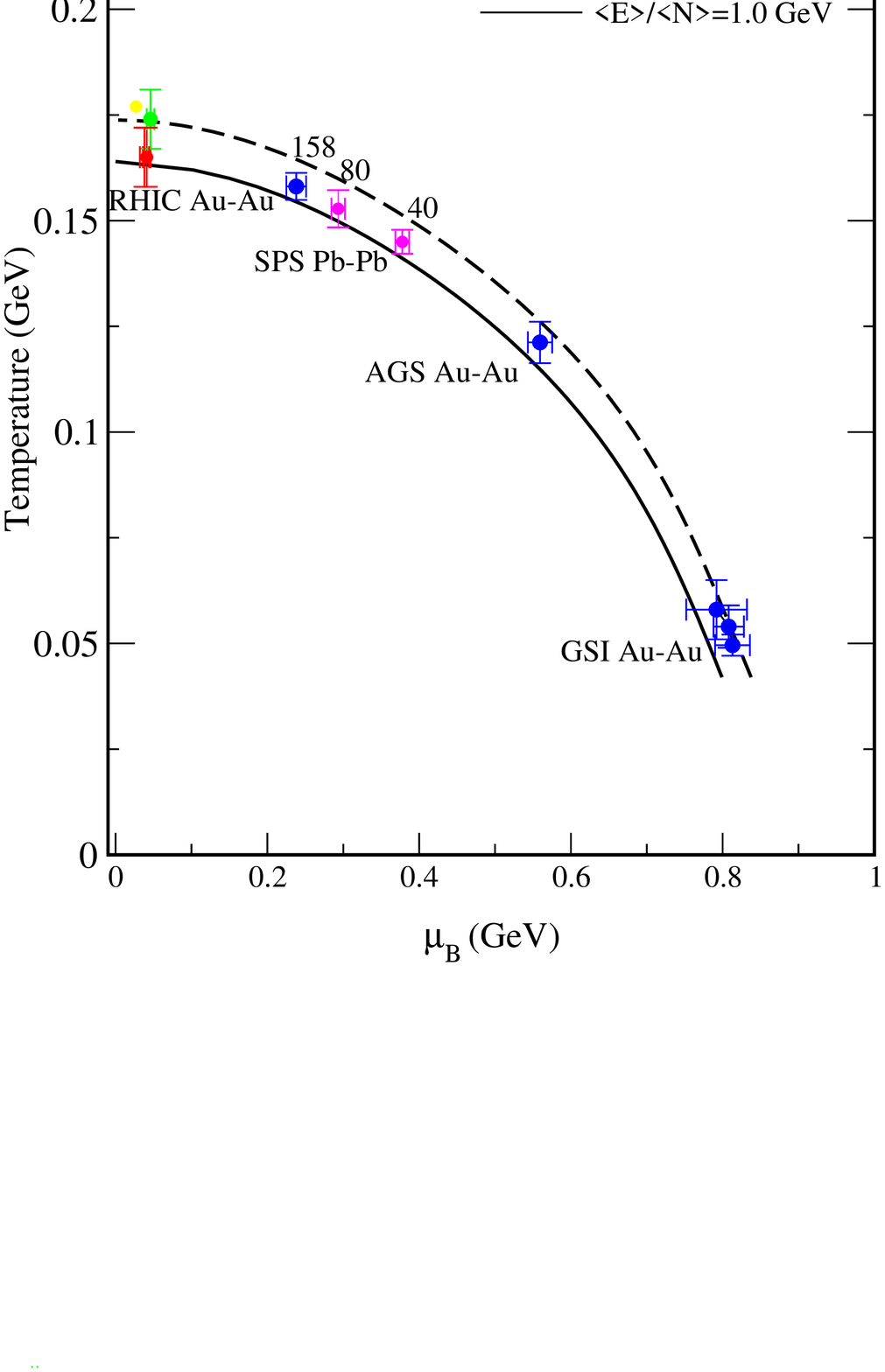}}
\vspace{-100pt}
\caption{
The chemical freeze-out points together with curves at fixed values of $E/N$.
}
\label{eovern}
\end{figure}

The thermal model has been extremely successful in bringing order to
a very large number of experimental results on particle yields in
relativistic heavy-ion collisions. The results for the temperature and 
baryon chemical potential have been found to be consistent with
 having $E/N = 1 $ GeV  
from the lowest beam energies up to the highest ones. This is illustrated in
Fig.~\ref{eovern}  which combines results from SIS up to RHIC.\\
The NA49 Collaboration  has recently  performed a series of measurements
of Pb-Pb collisions at 20, 30, 40, 80 and 158 AGeV beam energies.
When these results are combined with measurements at lower beam energies
from the
AGS and SIS they
reveal an unusually sharp variation with beam energy 
in the $\Lambda/\left<\pi\right>$, 
with
$\left<\pi\right>\equiv 3/2(\pi^++\pi^-)$,
 and in the
$K^+/\pi^+$ ratios. Such a strong variation with
energy does not occur in pp collisions and therefore indicates a
major difference in heavy-ion collisions. This transition
 has been referred to in Ref.~\cite{Gazdzicki} as the ``horn''.
A strong variation with energy of the $\Lambda/\left<\pi\right>$
ratio has been predicted on the basis of  arguments put forward
in~\cite{Gorenstein}. 
In~\cite{horn} 
another, less spectacular,
 possibility for the origin of the sharp maximum, namely as being due
to the transition from a baryon-dominated to a meson-dominated
hadronic gas has been suggested; 
the distinction being based on whether  the entropy of the hadronic gas
is dominated by baryons or by mesons.
For this purpose  various quantities
along the freeze-out curve~\cite{1gev} as a function of
$\sqrt{s_{NN}}$ have been studied in~\cite{horn}.

\begin{figure}[ht]
\centerline{
\epsfysize=32pc
\epsfxsize=24pc
\vspace{-145pt}
\epsfbox{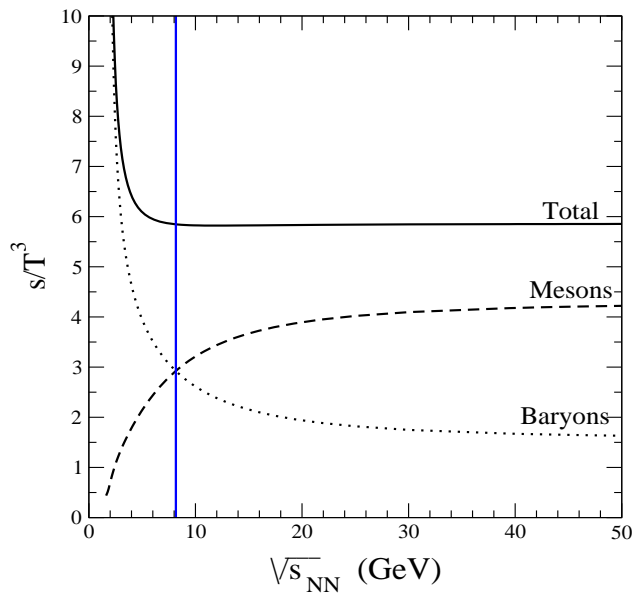}}
 \caption{The entropy density normalised to $T^3$
as a function of the beam energy as calculated in the thermal model using
THERMUS~\cite{thermus}.}
\label{entropy_s}
\end{figure}

In the thermal model a steep rise at low energies and a
subsequent flattening off leading to a mild maximum in the
$K^+/\pi^+$ ratio, was predicted many years
ago~\cite{cor,Becattini,max}. The sharpness of the observed 
peak therefore comes as a surprise. On the other hand, a sharp
peak in the $\Lambda/\left<\pi\right>$ ratio was predicted by the
thermal model~\cite{max} and is in good agreement with the
data. While the thermal model cannot explain the sharpness of
the peak in the $K^+/\pi^+$ ratio, there are nevertheless several
phenomena, giving rise to the rapid change, which warrant a closer
look at the model. 
%
%
To get a better estimate of the thermal parameters 
we have analyzed the entropy
density as a function of beam energy following the freeze-out
curve given in \cite{1gev} (see Fig. 2). 
There is a clear  transition from a  meson to a baryon-dominated
hadronic gas at
$\sqrt{s_{NN}}$ = 8.2 GeV. Above this value the  entropy is
carried mainly by mesonic degrees of freedom. It is remarkable that
the entropy density divided by $T^3$ is constant over the 
entire freeze-out curve, except for the  low-energy, SIS, energy region.
The line denoting the transition from 
a baryon-dominated to a meson-dominated hadron
gas  is shown in Fig.~\ref{entropy_s}. This line crosses
the freeze-out curve at a temperature of $T = 140$ MeV, when the
baryon chemical potential equals $\mu_B$ = 410 MeV. The
corresponding invariant energy is $\sqrt{s_{NN}}$ = 8.2 GeV.
%
%
%
\begin{figure}[ht]
\centerline{
\epsfysize=32pc
\epsfxsize=24pc
\vspace{-100pt}
\epsfbox{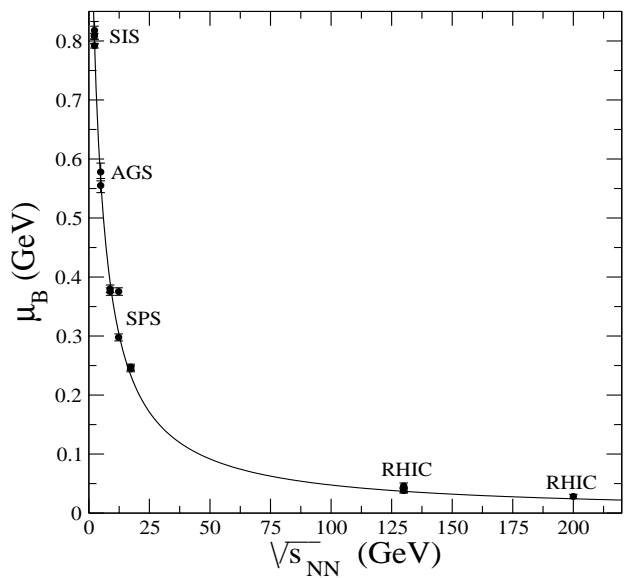}}
\caption{
The energy  dependence of the baryon chemical potential at freeze-out.
}
\label{mub_e}
\end{figure}
The strong decrease in the net baryon density 
is due to the fact that low energies are
characterized by a very low multiplicity of mesons and,
correspondingly, a very large baryon-to-meson ratio. As a
consequence, the baryon chemical potential is also very large.
 As the beam energy is increased, meson production increases
and the baryon chemical potential decreases. The
number of strange baryons produced in heavy-ion collisions at
different collision energies will follow the net baryon density
since a large baryon chemical potential will also enhance the
number of hyperons. 
As is well-known~\cite{review,manninen1}, the
thermal model description leads to a mild maximum in the
 $K^+/\pi^+$ ratio
which does not reproduce the so-called ``horn'' observed by
the NA49 collaboration~\cite{Gazdzicki}. 
 The observed
deviations at the highest SPS energy have been interpreted as a
lack of full chemical equilibrium in the strangeness sector,
leading to a strangeness suppression factor, $\gamma_s$, deviating
from its equilibrium value by about thirty percent. Detailed fits
using the thermal model in the region of the ``horn'' show
rapid variations in $\gamma_s$ \cite{manninen1} which do not lend 
themselves to any interpretation. There is no corresponding peak in 
the $K^-/\pi^-$ ratio because the production of $K^-$ is not tied 
to that of baryons. As the relative number of baryons
decreases with increasing energy, there is no corresponding decrease in the
number of $K^-$ as is the case with $K^+$ as these must be balanced by strange baryons.

It is worth noting that the maxima in the
 ratios for multi-strange baryons occur at ever
higher beam energies. 
 The higher the strangeness content of the baryon, the higher
in energy is the maximum. This behavior is due to a combination
of the facts that strangeness has to be balanced, the baryon
chemical potential decreases rapidly and the multi-strange baryons
have successively higher thresholds. The dependence of the baryon chemical potential on the beam energy is shown in Fig.~\ref{mub_e}.
It is to be expected that if these maxima do not all occur at the
same temperature, i.e. at the same beam energy, then the case for
a phase transition is not very strong. 

 In conclusion, while  the thermal model cannot explain
the sharpness of the peak in the $K^+/\pi^+$ ratio, its position
corresponds  precisely to a transition from a baryon-dominated to
a meson-dominated hadronic gas. This transition occurs at a
temperature $T = $ 140 MeV, a baryon chemical potential $\mu_B = $
410 MeV and an energy $\sqrt{s_{NN}} = $ 8.2 GeV. In the
thermal model this transition leads to a sharp peak in the
$\Lambda/\left<\pi\right>$ ratio, and to
 moderate peaks in the $K^+/\pi^+$, $\Xi^-/\pi^+$ and
$\Omega^-/\pi^+$ ratios. Furthermore, these peaks are at 
different energies in the thermal model. The thermal model
predicts that the maxima in the $\Lambda/\left<\pi\right>$,
$\Xi^-/\pi^+$ and $\Omega^-/\pi^+$ occur at increasing
beam energies. 

If the change in properties 
of the above excitation functions were associated with
a genuine deconfinement phase transition one would expect these
changes to  occur at the same beam energy.  It is clear that more data
are needed to clarify the precise nature of the sharp variation
observed by the NA49 collaboration.
\section*{References}
 
%
\end{document}